\documentclass[american,twoside]{article}
\usepackage{a4wide}
\usepackage{mathrsfs}
\usepackage{amsmath,amsfonts,amssymb,amsthm}
\usepackage{fancyhdr}
\usepackage{graphicx}

\usepackage{graphicx}
\usepackage{mathrsfs}
\usepackage{amsmath,amssymb,amsthm}

\usepackage{color}

\newtheorem{theorem}{Theorem}[section]
\newtheorem{definition}[theorem]{Definition}

\newtheorem{lemma}[theorem]{Lemma}

\theoremstyle{definition}

\def\bel{\begin{lemma}}
\def\eel{\end{lemma}}
\def\bet{\begin{theoreme}}
\def\eet{\end{theoreme}}
\def\bed{\begin{definition}}
\def\eed{\end{definition}}
\def \beq{\begin{equation}}
\def \eeq{\end{equation}}

\def \beq{\begin{equation}}
\def \eeq{\end{equation}}

\def\and {{\rm \; and \;}}

\newcommand{\R}{{\mathbb R}}

\def\cS{{\cal S}}
\def\cR{{\cal R}}

\def\fh{{\mathfrak h}}

\def\rr{{\mathbb R}}

\def\e{{\rm e}}
\def\i{{\rm i}}

\def\b0{{\boldsymbol{0}}}

\begin{document}

\title{A Dyson equation for non-equilibrium Green's functions in the partition-free setting}
%\titlerunning{Dyson equation for NEGF's}

\author{\sc H.D. Cornean$^{1}$, V. Moldoveanu$^{2}$, C.-A. Pillet$^3$
\\ \\ \\
$^1$Department of Mathematical Sciences\\
Aalborg University\\
Fredrik Bajers Vej 7G, 9220 Aalborg, Denmark
\\ \\
$^2$National Institute of Materials Physics\\
P.O. Box MG-7  Bucharest-Magurele, Romania
\\ \\
$^3$
Aix Marseille Univ, Universit\'e de Toulon, CNRS, CPT, Marseille, France
}
\maketitle
\thispagestyle{empty}

%\received{XXXX, revised XXXX, accepted XXXX} % do not change, will be filled in by the publisher
%\published{XXXX} % do not change, will be filled in by the publisher

% Please select about four verbal keywords for your manuscript.
%\keywords{Non-equilibrium Green Functions, Dyson equation, Partition free states}
\begin{quote}
{\bf Abstract.} We consider a small interacting sample coupled to several non-interacting leads. Initially, the system is at 
thermal equilibrium. At some instant $t_0$ the system is set into the so called partition-free transport scenario by 
turning on a bias on the leads. Using the theory of Volterra operators we rigorously formulate a Dyson equation for the
retarded Green's function and we establish a closed formula for the associated proper interaction self-energy. 
\end{quote}
%\pacs{73.21.La, 71.35.Cc, 03.67.Lx}

\maketitle

\section{Introduction}

The backbone of many-body perturbation theory (MBPT) is the interaction self-energy $\Sigma$ which appears in the Dyson 
equation for equilibrium or non-equilibrium  Green's function (NEGF). At equilibrium, the structure of $\Sigma$ is guessed by 
systematically using Wick's theorem and by analysing the resulting expansion into Feynman diagrams \cite{FW}. Approximation schemes 
(e.g.\;mean-field approach or RPA) correspond to partial resummation of series of diagrams contributing to $\Sigma $. 

In the finite temperature non-equilibrium regime of interacting systems the initial state cannot be in general 
connected to a non-interacting state in the remote past \cite{Adiabatic} and writing down statistical averages of time-dependent 
observables becomes cumbersome. The remedy for these technical difficulties is to 
combine the chronological $T$ and anti-chronological $\overline T$ time-ordering operators into a single operator ${\cal T}_C$ 
which allows an unambiguous book-keeping of time arguments on the two-branch Schwinger-Keldysh contour \cite{Sc,Ke,Danielewicz1}. 
This construction comes with a price: the non-equilibrium GFs turn to contour-ordered quantities as well and the various 
identities among them are not easy to recover. At a formal level one assumes the existence of a well-defined self-energy and 
then the contour-ordered Dyson equation splits via the Langreth rules \cite{La} into the Keldysh equation for the lesser/greater 
GFs and the Dyson equation for the retarded/advanced GFs (see the textbook \cite{HJ}). 

The existence of a self-energy for the contour-ordered GF is argued by the {\it formal} analogy between equilibrium and 
non-equilibrium quantum averages. Then a complete interaction self-energy can be defined \cite{Eq3.16}. 
In more recent formulations \cite{SvL} one starts from the differential equations of motions relating higher $n$-particle 
Green-Keldysh functions and then {\it truncates} the so-called Martin-Schwinger hierarchy \cite{Bonitz1} 
to identify various approximate interaction self-energies.  
      
Nowadays, the NEGFs formalism has grown up as a remarkable machinery, being extensively used for modelling quantum transport 
 in mesoscopic systems \cite{Myohanen}, molecules \cite{DvL} or even nuclear reactions \cite{Danielewicz2}. Nonetheless, 
some fundamental theoretical questions were only recently answered by fully exploiting the mathematical structure of the theory and {\it without} making 
any approximations. We refer here to: (i) the existence of non-equilibrium steady-state (NESS) in interacting open systems 
and (ii) the independence of the steady-state quantities from the initial state of the sample \cite{FMU,AJPP1,JOP3,CMP1,CMP2} both in 
the partitioning \cite{Ca1} and partition free \cite{Ci,HDP} settings. We recall here that in the partitioned case the system and
the biased leads are initially decoupled. 

In our recent work \cite{AHP} the NEGF formalism for open systems in the partitioning transport setting was rigorously 
treated in great detail and generality. In particular, we derived the Jauho-Wingreen-Meir formula (JWM) \cite{JWM} 
for the time-dependent current through an interacting sample by using only real-time quantities. 

In this short note we are interested in the partition-free regime which was adapted for interacting systems by 
Stefanucci and Almbladh \cite{SA}. Recently, the long-time limit of the energy current in the partition-free setting was discussed 
in Ref.\cite{Eich1} and the transient heat currents due to a temperature gradient were calculated in \cite{Eich2}.
 We briefly outline a rigorous formulation 
of the non-equilibrium Dyson equation for the retarded Green's function. Mathematical details are kept to a minimum while 
focusing on the  explicit construction of a complete interaction self-energy.       

The content of the paper goes as follows: the model and the notations are introduced in Section \ref{section2}, the main result and its proof are
 given in Section \ref{section3} while Section \ref{section4} is left for conclusions.

\section{Setting and notation.}\label{section2}
\subsection{Configuration space and Hamiltonians.}

We assume that a small sample is coupled to $M$ leads. The one-particle Hilbert space is of tight-binding type and can be written as $\mathfrak{h}=\mathfrak{h}_\cS\oplus \mathfrak{h}_\cR$ where $\mathfrak{h}_\cS$ is finite dimensional and $\mathfrak{h}_\cR=\oplus_{\nu=1}^M\mathfrak{h}_\nu$ describes the (finite or not) leads. Particles can only interact in the sample. One-particle operators are denoted with lower-case letters and their 
second quantized versions will be labeled by capital letters. The one-particle Hamiltonian of the decoupled system 
acquires a block-diagonal structure $h_{\rm D}=h_\cS\oplus h_\cR$ where $h_\cR=\oplus_{\nu=1}^Mh_\nu$ is supposed to be bounded. The lead-sample 
tunnelling Hamiltonian is defined as:
\begin{equation}\label{april1}
h_{\rm T}=\sum_{\nu=1}^M d_\nu\big (|f_\nu\rangle\langle g_\nu|
+|g_\nu\rangle\langle f_\nu|\big ),
\end{equation}
where $\nu$ counts the particle reservoirs, $f_\nu\in\fh_\nu$ and $g_\nu\in\fh_\cS$ are unit vectors and 
$d_\nu\in\rr$ are coupling constants. The one-particle Hamiltonian of the fully coupled system is then 
$h=h_{\rm D}+h_{\rm T}$.

We summarize below some useful identities from the second quantization machinery (see e.g. \cite{MR}). The total Fock space admits a factorization $\mathcal{F}=\mathcal{F}_\mathcal{S}\otimes \mathcal{F}_\mathcal{R}$.  By $a^{\#}(f)$ we mean either the creation operator $a^*(f)$ or the annihilation operator $a(f)$. We have $a^*(\lambda f)=\lambda a^*(f)$ and $a(\lambda f)=\overline\lambda a(f)$. 
The general form of the canonical anticommutation relations is:
\beq
\{a(f),a^\ast(g)\}=\langle f|g\rangle,\qquad\{a(f),a(g)\}=0.
\label{CAR}
\eeq
Here $\langle f|g\rangle$ denotes the scalar product in $\mathfrak{h}$.  Also, $a^{\#}(f)$ is bounded on the Fock space and $\|a^{\#}(f)\|\leq \|f\|$.

The interacting, coupled system, and with a potential bias $v_\nu$ on lead $\nu$ is described by:
\begin{equation}
K_v:=H+\sum_{\nu=1}^Mv_\nu N_\nu+\xi W,
\label{H_int}
\end{equation}
where $N_\nu$ is the particle number operator on lead $\nu$ (i.e., the second quantization of the orthogonal projection 
onto $\mathfrak{h}_\nu$), $v:=(v_1,\dots,v_M)\in\R^M$ is the bias vector and $$W=\frac{1}{2}\sum_{x,y\in\mathcal{S}}w(x,y) a^*(|x\rangle)a^*(|y\rangle)a(|y\rangle)a(|x\rangle)$$ is the second quantization of a two-body 
potential satisfying $w(x,y)=w(y,x)$ and $w(x,x)=0$ for all $x,y\in\cS$. Here $\xi\in\R$ stands for the interaction strength.

Assume that the bias is turned on at time $t=0$. Then the Heisenberg evolution of an observable $A$ at $t>0$ is
\begin{equation}\label{hc0}
\tau_{K_v}^t(A):=\e^{\i tK_v}A\e^{-\i tK_v},\quad t>0.
\end{equation}

If $h$ is a single-particle Hamiltonian, the associated Heisenberg evolution obeys:
\begin{equation}\label{Q1}
\tau_{H}^t(a^\#(f)):= \e^{\i t H}a^\#(f)\e^{-\i t H}=a^\#(\e^{\i th}f),
\end{equation}
and one has
\beq\label{hc0'}
[H,a^{*}(f)]=a^{*}(hf),\quad [H,a(f)]=-a(hf).
\eeq

Along the proof of the Dyson equation we shall encounter the operators:
\begin{equation}\label{W-comm}
b(f):=\i\xi[W,a (f)],\quad b^\ast(f):=\i\xi [W,a^\ast(f)].
\end{equation}
These operators vanish if $f$ is supported in the leads.

\subsection{The partition-free initial state.}

The initial state in the partition-free case is a Gibbs state characterized by the inverse temperature $\beta>0$ and the chemical potential $\mu\in\R$. It is given by the thermodynamic (i.e., infinite leads) limit of the density operator $\rho_{\rm pf}=Z^{-1}\e^{-\beta(K_0-\mu N)}$ where $Z={\rm Tr}_\mathcal{F}\;\e^{-\beta(K_0-\mu N)}$. In what follows we briefly explain how it is constructed. 

The interacting but decoupled and unbiased Hamiltonian is denoted by:
\begin{equation*}
K_D:=H_{\mathcal{S}}+\xi W +H_{\mathcal{R}}=K_0-H_T. 
\end{equation*}
The thermodynamic limit of $\rho_D=Z_D^{-1}\e^{-\beta(K_D-\mu N)}$
where $Z_D={\rm Tr}_\mathcal{F}\;\e^{-\beta(K_D-\mu N)}$ is a tensor product between a many-body Gibbs state 
$$\rho_S=\frac{1}{{\rm Tr}_{\mathcal{F}_{\mathcal{S}}}\e^{-\beta(H_\mathcal{S}+\xi W-\mu N_\mathcal{S})}}\e^{-\beta(H_\mathcal{S}+\xi W-\mu N_\mathcal{S})}$$ only acting on the finite dimensional Fock space $\mathcal{F}_{\mathcal{S}}$, and $M$ non-interacting $(\beta,\mu)$ Fermi-Dirac quasi-free states acting on each lead separately, where expectations can be computed with the usual Wick theorem. This special factorized initial state is denoted by $\big \langle \cdot \big\rangle _{\beta,\mu}$. For example, the expectation of a factorized observable of the type $\mathcal{O}=\mathcal{O}_\mathcal{S}\prod_{\nu=1}^M a^*(\tilde{f}_\nu)a(f_\nu)$ where $\tilde{f}_\nu,f_\nu\in \mathfrak{h}_\nu$ is:
$$\big \langle \mathcal{O} \big\rangle_{\beta,\mu}={\rm Tr}_{\mathcal{F}_{\mathcal{S}}}(\rho_\mathcal{S} \mathcal{O}_\mathcal{S})\prod_{\nu=1}^M\langle f_\nu|({\rm Id}+\e^{\beta(h_\nu-\mu)})^{-1}\tilde{f}_\nu\rangle.$$
 
Its connection with the partition-free state is as follows. Consider the operator $B(\alpha):=\e^{-i\alpha K_D}H_T\e^{i\alpha K_D}$, $\alpha\in\R$. 
From \eqref{april1} and using \eqref{Q1} we see that a generic term entering $B(\alpha)$ is 
$$\sum_\nu d_\nu \;  a^*(\e^{-\i\alpha h_\nu}f_\nu)\; \tau_{H_\mathcal{S}+\xi W}^{-\alpha}\big (a(g_\nu)\big ).$$
Since $h_\nu$ is bounded and $\mathcal{F}_\mathcal{S}$ is finite dimensional, this expression remains bounded for all complex values of $\alpha$. Then, the 
initial value problem  $$\Gamma'(x)= B(ix)\Gamma(x),\qquad\Gamma(0)={\rm Id},$$ has a unique solution given by a norm convergent Picard/Dyson/Duhamel iteration, 
with terms containing products of operators either living in the sample or in the leads. Before the thermodynamic limit, the operators 
$\Gamma(\beta)$ and $e^{\beta K_D}e^{-\beta K_0}$ satisfy the same differential equation and obey the same initial condition at $\beta=0$, 
hence they must coincide. 
Consequently, writing $\e^{-\beta K_0}=\e^{-\beta K_D} \Gamma(\beta)$ we obtain an appropriate expression for the thermodynamic limit: 
$\mathcal{O}$ being an arbitrary bounded physical observable, we have
\begin{align}\label{hc1}
\big\langle \mathcal{O}\big\rangle_{\rm pf }=\frac{\big\langle\Gamma(\beta)\mathcal{O}\big\rangle_{\beta,\mu}}{\big\langle\Gamma(\beta)\big\rangle_{\beta,\mu}}.
\end{align}

\subsection{Function spaces and Volterra operators.} Let $0<T<\infty$ be fixed and let $C_0^1([0,T];\mathfrak{h})$ be the space consisting of time dependent vectors $\phi(t)\in \mathfrak{h}$, $0\leq t\leq T$, which are continuously differentiable with respect to $t$, and $\phi(0)=0$. We also define $C([0,T];\mathfrak{h})$ to be the space of vectors which are only continuous in $t$, with no additional condition at $t=0$. We note that $C([0,T];\mathfrak{h})$ is a Banach space if we introduce the norm
$$|||\psi|||:=\sup_{0\leq t\leq T}\|\psi(t)\|_{\mathfrak{h}}.$$ 

 We say that an operator $A$ which maps $C([0,T];\mathfrak{h})$ into itself is a Volterra operator if there exists a constant $C_A<\infty$ such that 
$$
\|(A\psi)(t)\|_{\mathfrak{h}}\leq  C_A\int_0^t \|\psi(t')\|_{\mathfrak{h}} dt',\quad 0\leq t\leq T.$$
By induction one can prove:
$$
\|(A^n\psi)(t)\|_{\mathfrak{h}}\leq  C_A\frac{(C_AT)^{n-1}}{(n-1)!}\int_0^t \|\psi(t')\|_{\mathfrak{h}} dt',\quad n\geq 1. $$
This implies: 
$$|||A^n\psi|||\leq \frac{(C_A T)^n}{(n-1)!}|||\psi|||$$
which leads to the conclusion that the operator norm of $A^n$ is bounded by $\frac{(C_A T)^n}{(n-1)!}$. 
In particular, the series $\sum_{n\geq 1}(-1)^nA^n$ converges in operator norm and defines a Volterra operator with a constant less than $C_A\e^{TC_A}$. Thus, $({\rm Id}+A)^{-1}={\rm Id}+\sum_{n\geq 1}(-1)^nA^n$  always exists and 
$A({\rm Id}+A)^{-1}$ is a Volterra operator. 

\subsection{Retarded NEGF's.}

Let $\{e_j\}$ be an arbitrary orthonormal basis in $\mathfrak{h}$. Define the map $G_0: C([0,T];\mathfrak{h})\mapsto C_0^1([0,T];\mathfrak{h})$ given by:
\begin{align}\label{hc2}
\langle e_j|(G_0\psi)(t)\rangle:=-\i \int_0^t \langle e_j|\e^{-\i(t-t')h_v}\psi(t')\rangle dt',
\end{align}
where $h_v$ denotes the single-particle Hamiltonian of the non-interacting coupled and biased system. 
One can check that $G_0$ is invertible and if $\phi\in C_0^1([0,T];\mathfrak{h})$:
\begin{align}\label{hc3}
(G_0^{-1}\phi)(t)=\i \partial_t\phi(t)-h_v\phi(t)\in C([0,T];\mathfrak{h}).
\end{align}
By definition, the retarded non-equilibrium Green operator in the partition-free setting $G_\xi:C([0,T];\mathfrak{h})\mapsto C_0^1([0,T];\mathfrak{h})$ is given by:
\begin{align}\label{hc4}
\langle e_j|(G_\xi\psi)(t)\rangle:=-\i \int_0^t \big \langle 
\{ \tau_{K_v}^{t'}(a^*(\psi(t'))), \tau_{K_v}^{t}(a(e_j))\}\big\rangle_{\rm pf}\;  dt'.
\end{align}
Using \eqref{Q1} and \eqref{CAR} we see that $G_\xi$ coincides with $G_0$ when $\xi=0$. One can show that
\begin{align}\label{hc5}
\|(G_\xi\psi)(t)\|_{\mathfrak{h}}\leq  2\int_0^t \|\psi(t')\|_{\mathfrak{h}} dt',
\end{align}
so that $G_\xi$ is a Volterra operator.
The integral kernel of $G_\xi$ is nothing but the more familiar retarded NEGF given by:
\begin{align}\label{hc4'}
G_\xi^R(e_j,t;e_m,t'):=-\i \theta(t-t') \big\langle \{ \tau_{K_v}^{t'}(a^*(e_m)), \tau_{K_v}^{t}(a(e_j))\}\big\rangle_{\rm pf}\;,
\end{align}
and 
\begin{align}\label{hc4''}
\langle e_j|(G_\xi\psi)(t)\rangle=\sum_m\int_0^t G_\xi^R(e_j,t;e_m,t')\langle e_m|\psi(t')\rangle   dt'.
\end{align}
The advanced NEGF can be defined as:
\begin{align*}
G_\xi^A(e_j,t;e_m,t'):=-G_\xi^R(e_j,t';e_m,t)=\i \theta(t'-t) \big\langle 
\{ \tau_{K_v}^{t'}(a^*(e_m)), \tau_{K_v}^{t}(a(e_j))\}\big\rangle_{\rm pf}\;.
\end{align*}
All properties of the advanced NEGF can be immediately read off from those of the retarded one.

\section{Irreducible self-energy and Dyson equation.}\label{section3}
Here is the main result of our paper. 
\begin{theorem}\label{thm1}
 The bounded linear map $\widetilde{\Sigma}_\xi$ defined on $C([0,T];\mathfrak{h})$  by
\begin{align}\label{hc9}
\langle e_j|(\widetilde{\Sigma}_\xi\phi)(t)\rangle:=-\i \int_0^t \big\langle\{ \tau_{K_v}^{t'}(b^*(\phi(t'))), 
\tau_{K_v}^{t}(b(e_j))\}\big\rangle_{\rm pf}\;  dt'+ \i \big\langle\tau_{K_v}^{t}(\{a^*(\phi(t)),b(e_j)\})\big\rangle_{\rm pf}\;,
\end{align}
obeys:
\begin{align}\label{hc6}
G_\xi=G_0+G_0\widetilde{\Sigma}_\xi G_0.
\end{align}
Moreover, the operator $G_0\widetilde{\Sigma}_\xi$ is a Volterra operator, the inverse $({\rm Id} +G_0\widetilde{\Sigma}_\xi)^{-1}$ exists, and by defining 
\begin{align}\label{hc20}
\Sigma_\xi:=\widetilde{\Sigma}_\xi\left ({\rm Id}+ G_0 \widetilde{\Sigma}_\xi\right )^{-1}
\end{align}
we have:
\begin{align}\label{hc21}
G_\xi=G_0+G_0\Sigma_\xi G_\xi.
\end{align}
Finally, $G_0 \Sigma_\xi$ is also a Volterra operator and
\begin{align}\label{hc30}
G_\xi=\left ({\rm Id}- G_0 \Sigma_\xi\right )^{-1}G_0.
\end{align}
As in the physical literature Eq.(\ref{hc20}) defines the irreducible self-energy operator $\Sigma_\xi$ in terms of 
the reducible part $ \widetilde{\Sigma}_\xi$. 
\end{theorem}

\vspace{0.5cm}

\subsection{Proof: step 1.} First we will show that the identity:
\begin{align}\label{hc7}
G_0^{-1}G_\xi={\rm Id}+F_\xi
\end{align}
holds on $C([0,T];\mathfrak{h})$, where the map $F_\xi$ is given by
\begin{align}\label{hc8}
\langle e_j|(F_\xi\psi)(t)\rangle:=\int_0^t \big\langle\{ \tau_{K_v}^{t'}(a^*(\psi(t'))), \tau_{K_v}^{t}(b(e_j))\}\big\rangle_{\rm pf}\;  dt'.
\end{align}
Using \eqref{hc3} and \eqref{hc4} we have: 
\begin{align}\nonumber
&\langle e_m|(G_0^{-1}G_\xi\psi)(t)\rangle =\langle e_m|\psi(t)\rangle-\sum_j \langle e_m|h_ve_j\rangle \langle e_j|(G_\xi\psi)(t)\rangle
\label{hc11}\\
&+\int_0^t \big\langle\{ \tau_{K_v}^{t'}(a^*(\psi(t'))), \partial_t\tau_{K_v}^{t}(a(e_m))\}\big\rangle_{\rm pf}\;  dt'.
\end{align}
From the antilinearity of the annihilation operators we get
\begin{align*}
\sum_j \langle e_m|h_ve_j\rangle \langle e_j|(G_\xi\psi)(t)\rangle=
-\i \int_0^t \big\langle\{ \tau_{K_v}^{t'}(a^*(\psi(t'))), \tau_{K_v}^{t}(a(h_ve_m))\}\big\rangle_{\rm pf}\;  dt'.
\end{align*}
Also, using \eqref{hc0}, \eqref{hc0'} and \eqref{W-comm} we obtain the identity:
$$\partial_t\tau_{K_v}^{t}(a(e_m))=-i\tau_{K_v}^{t}(a(h_ve_m))+\tau_{K_v}^{t}(b(e_m)).$$
After introducing the last two identities into \eqref{hc11} we see that two terms cancel each other and we obtain \eqref{hc8}. 

\subsection{Proof: step 2.} The second step consists of showing that $F_\xi$ can be written as $\widetilde{\Sigma}_\xi G_0$, with $\widetilde{\Sigma}_\xi$ as in \eqref{hc9}. In order to identify $\widetilde{\Sigma}_\xi$ we compute for every $\phi\in C_0^1([0,T];\mathfrak{h})$ the quantity (remember that $a^*$ is linear): 
\begin{align}\nonumber
\langle e_j|(F_\xi G_0^{-1}\phi)(t)\rangle &=\i\int_0^t \big\langle\{ \tau_{K_v}^{t'}(a^*(\partial_{t'}\phi(t'))), \tau_{K_v}^{t}(b(e_j))\}\big\rangle_{\rm pf}\;  dt'\label{hc12}\\
&-\int_0^t \big\langle\{ \tau_{K_v}^{t'}(a^*(h_v\phi(t'))), \tau_{K_v}^{t}(b(e_j))\}\big\rangle_{\rm pf}\;  dt'.
\end{align}
Another key identity is: 
\begin{align*}
\tau_{K_v}^{t'}(a^*(\partial_{t'}\phi(t')))=\partial_{t'}\left (\tau_{K_v}^{t'}(a^*(\phi(t')))\right )-\i \tau_{K_v}^{t'}(a^*(h_v\phi(t')))-\tau_{K_v}^{t'}(b^*(h_v\phi(t'))).
\end{align*}
Inserting this identity in \eqref{hc12}, integrating by parts with respect to $t'$ and using that $\phi(0)=0$, we obtain \eqref{hc9}. 

\subsection{Proof: step 3.} From the first two steps we derive \eqref{hc6}. 
From \eqref{hc9} and \eqref{hc5} we see that $A=G_0 \widetilde{\Sigma}_\xi$ is a Volterra operator for which there exists a $T$-dependent constant $C<\infty$ such that 
\begin{align}\label{hc10}
\|(A\psi)(t)\|_{\mathfrak{h}}\leq  C\; \int_0^t \|\psi(t')\|_{\mathfrak{h}} dt',\quad 0\leq t\leq T.
\end{align}
Then $({\rm Id} +A)^{-1}$ exists and it is given by a norm convergent Neumann series $\sum_{n\geq 0}(-1)^nA^n$, as long as $T<\infty$. 
 We write 
$$G_0=\left ({\rm Id}+ G_0 \widetilde{\Sigma}_\xi\right )^{-1}G_\xi$$
and we can choose $\Sigma_\xi$ as in \eqref{hc20}, which finishes the construction of the proper self-energy. 

\subsection{Consequences.} We list a few remarks concerning our main theorem.

\noindent (i) The integral kernel  of $\widetilde{\Sigma}_\xi$ (see \eqref{hc9}) is given by 
\begin{align}\label{hc22}
\widetilde{\Sigma}_\xi^R(e_j,t;e_m,t'):=-\i \theta(t-t') \big\langle\{ \tau_{K_v}^{t'}(b^*(e_m)), \tau_{K_v}^{t}(b(e_j))\}\big\rangle_{\rm pf}
+ \i \delta(t-t')\big\langle\tau_{K_v}^{t}(\{a^*(e_m),b(e_j)\})\big\rangle_{\rm pf}\;.\nonumber
\end{align}
If either $e_j$ or $e_m$ belongs to the leads, then the above matrix element equals zero. The explanation for the first term is that at least one of the two operators $b(e_j)$ and $b^*(e_m)$ defined through \eqref{W-comm} would be zero in this case, because the self-interaction $W$ 
is only supported in the sample, hence it commutes with any observable supported on the leads. For the second term, assume that $e_j$ is from the sample while $e_m$ is from the leads. Then since $b(e_j)$ is a sum of products of three creation/annihilation operators from the sample, it anticommutes with $a^*(e_m)$. 

The proper self-energy $\Sigma_\xi$ has the same support property. One recognizes that $\widetilde{\Sigma}_\xi^R(e_j,t;e_m,t')$ is a
 reducible self-energy . In the diagrammatic language all terms contributing to $\widetilde{\Sigma}_\xi^R(e_j,t;e_m,t')$ connect 
to other diagrams by incoming and outgoing $G_0$-lines.

\noindent (ii)
If both $e_j=x$ and $e_m=y$ are located in the small sample, then from \eqref{hc21} we see that in order to compute $G_\xi^R(x,t;y,t')$ we only need to know the values of $G_0$ restricted to the small sample (besides $\Sigma_\xi$, of course). From \eqref{hc2} we have: 

\begin{align*}
&G_0^R(x,t;y,t')=-\i \theta(t-t') \langle x|\e^{-\i(t-t')h_v}y\rangle ,
\end{align*}
with $x,y\in\mathcal{S}$. Such matrix elements can be computed from the resolvent $(h_v-z)^{-1}$ restricted to the small sample; we note that via the Feshbach formula, the biased leads appear as a non-local ``dressing'' potential which perturbs $h_\mathcal{S}$, see \cite{CMP1} for details.

At the level of integral kernels, the Dyson equation \eqref{hc21} reads as:    
\begin{align*}
G_\xi^R(x,t;y,t')=G_0^R(x,t;y,t')+\sum_{u,v\in \mathcal{S}}\int_0^t ds\int_0^s ds' G_0^R(x,t;u,s)\Sigma_\xi^R (u,s;v,s')G_\xi^R(v,s';y,t').
\end{align*}

\noindent (iii) Assume that we can write $\Sigma_\xi$ as $\Sigma_{\rm app}+\Sigma'$, where $\Sigma_{\rm app}$ is an approximating Volterra operator. If $G_{\rm app}=({\rm Id}-G_0\Sigma_{\rm app})^{-1}G_0$ is the solution of the approximate Dyson equation $G_{\rm app}=G_0+G_0\Sigma_{\rm app}G_{\rm app}$, then we have: 

$$G_\xi=G_{\rm app}+G_{\rm app}\Sigma'G_\xi$$
and $G_\xi=({\rm Id}-G_{\rm app}\Sigma')^{-1}G_{\rm app}$.

\noindent (iv) The limit $T\to\infty$ is a difficult problem. To the best of our knowledge, the only rigorous mathematical results concerning the existence of a steady-state regime in partition free-systems are \cite{CMP1,CMP2}. Under certain non-resonant conditions and for $\xi$ small enough, one can prove that  a quantity like $G_\xi^R(e_m,t'+s;e_n,t')$, where $s>0$ is fixed, will have a limit as $t'\to\infty$. This is definitely not guaranteed to happen in all cases, not even in non-interacting systems, due to bound states which may produce persistent oscillations.

(v) One may generalize the present setting in order to allow a non-trivial time dependence of the bias, the only difference 
would appear in the evolution groups which now would have time-dependent generators. Also, the notation and formulas would be more involved, but no new mathematical issues would appear.

\section{Conclusions}\label{section4}

We presented a non-perturbative approach to the partition-free transport problem. Starting from the Volterra operator associated to the retarded 
Green's function we establish its Dyson equation, and we derive closed formulas for the reducible and irreducible self-energies. 
The proof is rigorous yet elementary in the sense that although the partition-free scenario is a genuine non-equilibrium regime we do not 
use contour-ordered operators. A Keldysh equation for the lesser Green's function should be established following the same lines of reasoning, with the extra difficulty induced by the fact that in the partition free setting, the small sample is not empty at $t=0$. 

Unravelling the connection between the closed formula \eqref{hc9} and the diagrammatic approach
remains an open problem. Although the anti-commutator structure $\big\langle\tau_{K_v}^{t}(\{a^*(\phi(t)),b(e_j)\})\big\rangle_{\rm pf} $  in Eq. \eqref{hc9}
looks less familiar one can speculate that the systematic application of the Wick theorem should eventually recover various classes of diagrams. 
A possible approximation in the self-energy would be to replace the interacting propagator $\tau_{K_v}^{t}(\cdot)$ with the non-interacting one $\tau_{H_v}^{t}(\cdot)$, 
where $K_v=H_v+\xi W$. Note however that the application of the Wick theorem is technically challenging due to the extra term $\Gamma(\beta)$ appearing in \eqref{hc1}.

Given the fact that the partition-free setting is less studied in the literature, yet more intuitive on physical grounds than the partitioned approach, 
we hope that our investigation will trigger more efforts from both the physical and mathematical-physics communities.
Our main message is that one can properly formulate some of the central equations of the 
many-body perturbation theory (MBPT) in a direct way, paying close attention to fundamental issues like convergence, 
existence, uniqueness, stability, and at the same time, trying to obtain precise error bounds for a given approximation 
of the self-energy. The Volterra theory guarantees that for relatively small $T$'s one can "keep doing what one has been 
doing"; however, the large time behavior like for example the existence of steady states and the speed of convergence seem 
to be very much dependent on the system and no general recipe can work out in all cases.

%\begin{acknowledgement}
\textbf{Acknowledgments.} V.M.\;acknowledges financial support by the CNCS-UEFISCDI Grant PN-III-P4-ID-PCE-2016-0084 and from the 
Romanian Core Research Programme PN16-480101. H.C.\;acknowledges financial 
support by Grant 4181-00042 of the Danish Council for Independent Research $|$ Natural Sciences. C.A.P.\;acknowledges 
financial support by the ANR, Grant NONSTOPS (ANR-17-CE40-0006),
%\end{acknowledgement}


\begin{thebibliography}{20}
\bibitem{Adiabatic} Such a connection would rely on the so called adiabatic assumption which however must be checked case 
by case (see Chapter 4.3 in Ref.\cite{SvL}).
\end{thebibliography}
\end{document}